\documentclass[
journal=nalefd,
manuscript=letter,
layout=twocolumn]{achemso}

\usepackage{chemformula} 
\usepackage[T1]{fontenc} 
\usepackage{graphicx}
\usepackage{lineno}
\usepackage{helvet}

\let\oldmaketitle\maketitle
\let\maketitle\relax

\author{Liam S. Farrar}
\email{L.S.Farrar@bath.ac.uk}
\affiliation[Bath]{Department of Physics, University of Bath, Bath BA2 7AY, United Kingdom}
\author{Aimee Nevill}
\affiliation[Bath]{Department of Physics, University of Bath, Bath BA2 7AY, United Kingdom}
\author{Zhen Jieh Lim}
\affiliation[Bath]{Department of Physics, University of Bath, Bath BA2 7AY, United Kingdom}
\author{Geetha Balakrishnan}
\affiliation[Warwick]{Department of Physics, University of Warwick, Coventry CV4 7AL, United Kingdom}
\author{Sara Dale}
\affiliation[Bath]{Department of Physics, University of Bath, Bath BA2 7AY, United Kingdom}
\author{Simon J. Bending}
\affiliation[Bath]{Department of Physics, University of Bath, Bath BA2 7AY, United Kingdom}

\title[ACS Nano SQUID]
  {Superconducting Quantum Interference in Twisted van der Waals Heterostructures}

\begin{document}

\twocolumn[
\begin{@twocolumnfalse}
\oldmaketitle
\begin{abstract}
Modern Superconducting QUantum Interference Devices (SQUIDs) are commonly fabricated from either Al or Nb electrodes, with an in-situ oxidation process to create a weak link between them.
 However, common problems of such planar nano- and micro-SQUIDs are hysteretic current-voltage curves, and a shallow flux modulation depth.
 Here, we demonstrate the formation of both Josephson junctions and SQUIDs using a dry transfer technique to stack and deterministically misalign flakes of NbSe$_{2}$; allowing one to overcome these issues.
 The Josephson dynamics of the resulting twisted NbSe$_{2}$-NbSe$_{2}$ junctions are found to be sensitive to the misalignment angle of the crystallographic axes.
 A single lithographic process was then implemented to shape the Josephson junction into a SQUID geometry with typical loop areas of $\simeq$ 25 $\mu m^{2}$ and weak links $\simeq$ 600 nm wide.
 These devices display large stable current and voltage modulation depths of up to $\Delta I_{c} \simeq$ 75$\%$ and $\Delta V \simeq$ 1.4 mV respectively.

\end{abstract}
\end{@twocolumnfalse}
]
Superconducting QUantum Interference Devices (SQUIDs) are key components in the development of ultra-sensitive electric and magnetic measurement systems.\cite{Fagaly2006} 
The basic SQUID design consists of a superconducting ring intersected by one (rf SQUID) or two (dc SQUID) Josephson junctions; the latter consisting of two superconducting electrodes coupled by weak links that allow the flow of supercurrent.
 The weak link can take a variety of forms including point contacts, physical constrictions, or heterostructures consisting of a thin normal metal or insulating barrier separating the two superconductors.\cite{Likharven1979} 
 The latter type is typically fabricated by deposition of metallic superconductors such as Al and Nb, with the tunnelling barrier formed by in-situ oxidation.\cite{Gurvitch1983}
 This oxide barrier varies in thickness on the atomic scale and often contains defect traps, which can lead to highly non-uniform supercurrent distributions.\cite{Zeng2015}
 In addition, over time oxygen atoms in the oxide barrier diffuse out, altering the normal state resistance of the junction resulting in a variation of the critical current, a process which can be detrimental to the longevity of the device.\cite{Pop2012}
 Since Josephson junctions are the basis of many superconducting technologies such as qubits,\cite{Kjaergaard2020} quantum metrology,\cite{Scherer2012} and superconducting quantum interference devices (SQUIDs), development and incorporation of new materials with improved properties and functionality is vital.

\begin{figure*}[t]
 \centering
  \includegraphics[trim={0cm 0cm 0cm 0cm}, width=1\linewidth,clip=true]{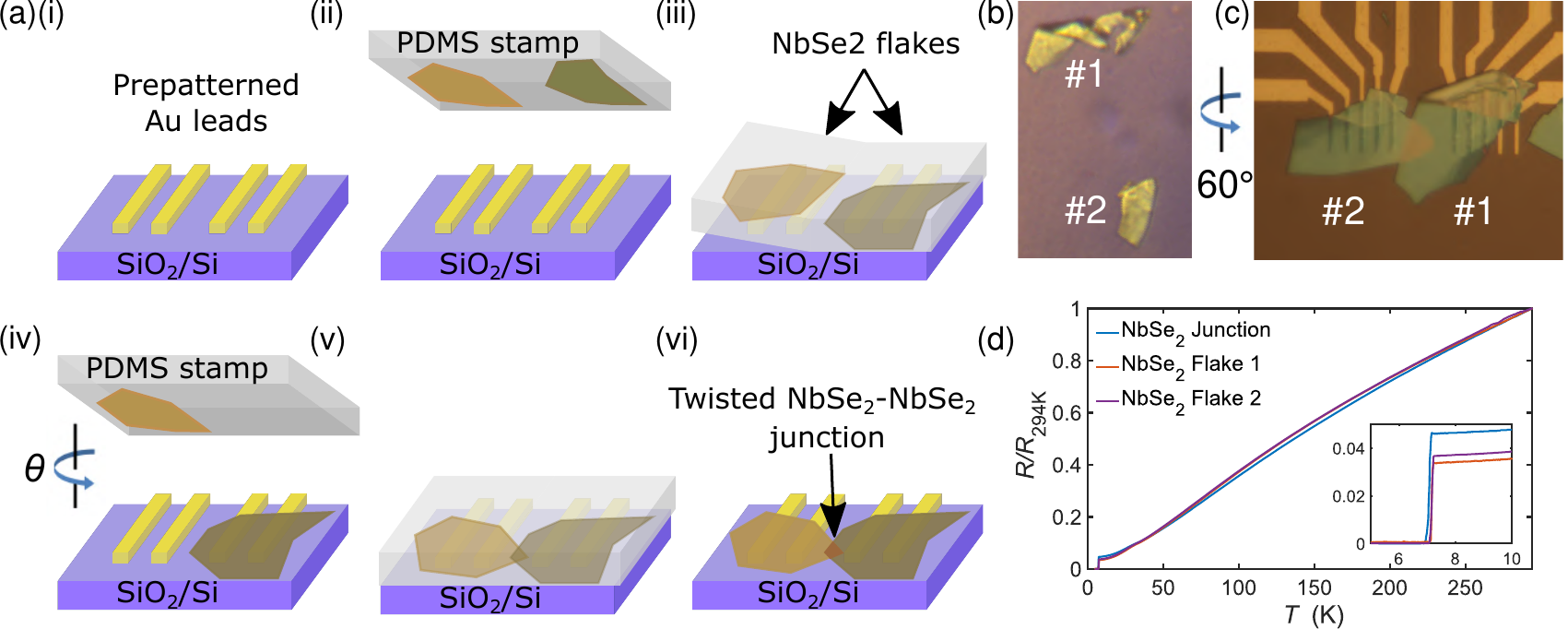}
  \caption{
a) Schematic of the device fabrication method. i) Au contacts are deposited onto a Si/SiO$_{2}$ substrate. ii) A single exfoliation is made of a bulk NbSe$_{2}$ crystals and the resulting flakes are transferred onto a PDMS stamp. iii) The PDMS is brought into contact with the substrate which is tilted at a small angle. The PDMS is slowly pressed into the substrate until the boundary of the PDMS-substrate contact region lies beyond one of the flakes. iv) The PDMS is then retracted, leaving the first flake deposited onto the Au contacts. The substrate is then rotated such that the crystallographic axes of the two NbSe$_{2}$ flakes are now misaligned by an angle $\theta$. v) The second flake is positioned above the first and the PDMS is brought into contact with the substrate. vi) The PDMS is now retracted leaving the second flake contacting both the first flake and the Au contacts. b) Optical micrograph of two NbSe$_{2}$ flakes on a PDMS stamp. c) Optical micrograph of the two overlapping NbSe$_{2}$ flakes on SiO$_{2}$/Si. Here one of the flakes (labelled $\#$2) has been rotated by 60$^{\circ}$. d) Temperature-dependence of the normalised resistance \textit{R}(\textit{T})/\textit{R}(294 K) for the two NbSe$_{2}$ flakes and the overlapping junction region between them.
   }

  \label{fig:Figure_1}
\end{figure*}

 An alternative route to the fabrication of oxide-free Josephson junctions is the stacking of two-dimensional materials (2D) into vertical van der Waals (vdW) heterostructures,\cite{Wang2013} a technique which holds promise for creating atomically clean and defect free interfaces.
 Additionally, vdW-based devices provide access to a new selection of superconducting materials such as Bi$_{2}$Sr$_{2}$CaCu$_{2}$O$_{8+\delta}$,\cite{Yu2019} FeSe,\cite{Farrar2020} and 2H-NbSe$_{2}$.\cite{xi2016}.
 These materials all share a common characteristic in that they have a layered structure and easily cleave perpendicular to the $c$-axis.
 These properties, along with the availability of dry transfer techniques,\cite{Castellanos201} has allowed the fabrication of vdW heterostructure devices comprised of two or more mechanically exfoliated flakes.
 Such devices include junctions with dielectric tunnelling barriers,\cite{Khestanova2018,Dvir2018} superconducting-normal-superconducting Josephson junctions,\cite{Kim2017} and van der Waals interface Josephson junctions,\cite{Yabuki2016} which may prove useful in the formation of superconducting qubits.\cite{Liu2019} 
 Furthermore, a vdW-based device offers an additional variable in the form of controlling the relative twist angle between the crystallographic axes of the two materials; a degree of freedom which is not available in conventional heterostructures.
 This twisting creates a misalignment between the two crystals, changing the atomic registry at the interface and leading to angular-dependent interlayer interactions.
 This enables one to tune the electronic coupling via the twist angle, leading to effects such as band hybridisation,\cite{Ohta2012,Thompson2020} minigaps and band replicas due to scattering on moiré potentials,\cite{Pierucci2016} charge transfer and changes in effective masses \cite{Yeh2016}, and has led to the discovery of a moiré superlattice in graphene/h-BN,\cite{Dean2013} as well as electrically tunable superconductivity in twisted bilayer graphene.\cite{Cao2018} 
 This new field of study named twistronics, may prove valuable in the incorporation of two-dimensional materials into more complex vdW superconducting devices such as qubits and SQUIDs. 
 
 \begin{figure}[t]
 \centering
  \includegraphics[trim={0cm 0cm 0cm 0cm}, width=1\linewidth,clip=true]{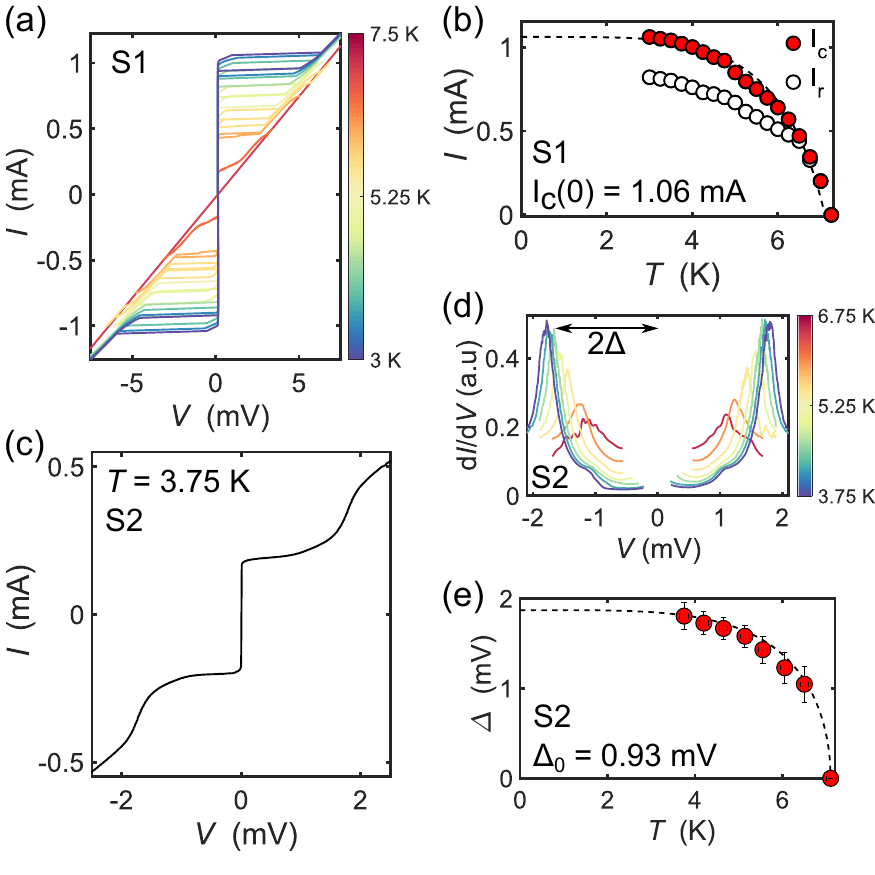}
   \caption{
(a) Current-voltage ($I$-$V$) characteristics of a NbSe$_{2}$-NbSe$_{2}$ junction (sample S1, $\theta$ $\approx$ 6$^{\circ}$) for temperatures between $T$ = 3 K and 7.5 K. b) Temperature dependence of the critical current $I_{\rm c}$ (red) and retrapping current (white). The dashed line is calculated from AB theory (see text). c) $I$-$V$ characteristics for sample S2 ($\theta$ $\approx$ 20$^{\circ}$) at $T$ = 3.75 K. d) Temperature dependence of d$I$/d$V$ versus V between $T$ = 3.75 and 6.75 K. e) Temperature dependence of the estimated superconducting gap $\Delta$. The dashed line is calculated from an interpolation approximation to BCS theory.
}
  \label{fig:Figure_3}
\end{figure}
 
In this study, twisted NbSe$_{2}$-NbSe$_{2}$ heterostructures are formed by stacking two NbSe$_{2}$ flake with a well-determined misalignment of the crystallographic axes using the procedure described in Methods and highlighted in Fig \ref{fig:Figure_1}a. 
 Briefly, two NbSe$_{2}$ are sequentially stacked with a small overlap onto a Si/SiO$_{2}$ substrate with pre-patterned Au contacts and encapsulated with a thin layer of hexagonal boron nitride. 
 Typical junction areas are on the order of 20-60 $\mu$m$^{2}$ (Fig. \ref{fig:Figure_1}c).
 
Before describing the characterisation of devices patterned with SQUID geometries, we investigate the transport properties of twisted NbSe$_{2}$-NbSe$_{2}$ junctions.
 The temperature-dependence of the normalised four-point resistance \textit{R}(\textit{T})/\textit{R}(294K) from 4 to 294 K is shown in Fig. \ref{fig:Figure_1}d for two NbSe$_{2}$ flakes with a relative twist angle of 60$^{\circ}$, as well as the resulting junction formed by the overlap region.
We note that there is an uncertainty of $\pm$1$^{\circ}$ in the misalignment angle of devices arising from both the resolution of the rotation stage and unwanted movement of the PDMS during the stamping progress.
Both flakes and junction show metallic transport behaviour which is phonon-limited at high temperature (\textit{R} $\propto$ \textit{T}) and disorder limited at low temperature  (\textit{R} approaches a constant value) before reaching the superconducting state at \textit{T}$_{c}$ $\sim$ 7 K.\cite{Naito1982}
As the $T_{c}$ of NbSe$_{2}$ is known to reduce in flakes $<$ 10 layers,\cite{Khestanova2018,Dvir2018} all flakes used were at least 10 nm, ensuring negligible suppression of $T_{\rm c}$.
 The residual resistance ratio (RRR), defined as \textit{R}(\textit{T}=294K)/\textit{R}(\textit{T}=8K), is found to be $\sim$ 30 for the two NbSe$_{2}$ flakes, with a small variation due to the slightly different flake thicknesses.
 This is in contrast to the RRR of the junction region which is found to vary between $\sim$10-25 depending on the twist angle.
 A close up of the superconducting transition is shown in Fig \ref{fig:Figure_1}(d), where it can be seen that the junction displays a slightly broader transition width when compared to the individual NbSe$_{2}$ flakes, indicative of additional disorder at the interface.

\subsubsection{Josephson junction performance}

Next,
 we examine the current-voltage (\textit{I}-\textit{V}) characteristics of a NbSe$_{2}$-NbSe$_{2}$ junction for temperatures $T$ = 3 to 7.5 K (S1, Fig. \ref{fig:Figure_3}a). 
 Here the misorientation angle of the junction is estimated to be $\theta \sim$ 6$^{\circ}$ from comparison of optical micrographs of the two flakes. 
 From the presence of distinct retrapping currents in the \textit{I}-\textit{V} curves, it is clear that the junction 
behaviour is typical of an underdamped Josephson junction.
 This indicates that the vdW interface decouples the superconducting order parameters and creates a weak link that allows the flow of a Josephson supercurrent.
 Each NbSe$_{2}$ flake has multiple electrical contacts, allowing the use of different voltage lead configurations to confirm that the bulk critical current of each flake is substantially larger than that of the junction.
Next, we quantify the junction dynamics by examining the McCumber parameter $\beta_{c}$, found by fitting the ratio between the retrapping current $I_{r}$ and the critical current $I_{c}$ and comparing it to a microscopic model.\cite{Likharev1986}
 $I_{r}/I_{c}$ is found to be $\sim$0.89 at $T$ = 3 K, leading to a McCumber parameter $\beta_{c}=$ 0.6, placing this device in the 'weakly underdamped' category.
 Using the relationship $\beta_{c} = 2eI_{c}R^{2}_{N}C\hbar^{-1}$ the junction capacitance, $C$, was found to be 1.6x$10^{-13}$F, yielding a specific capacitance of $C/A$ = 0.8 $\mu Fcm^{-2}$.
 Assuming the permittivity is that of vacuum, this leads to a gap between the two flakes of $<$1 nm, comparable to that of the $c$-axis period in 2H-NbSe$_{2}$ of 0.63 nm.
 
The temperature dependence of the critical current, $I_{\rm c}$, can be calculated using Ambegaokar-Baratoff (AB) theory\cite{Barone1982}, in which $I_{\rm c}(T)$ is expressed as
\begin{equation} \label{eq:AB}
\frac{I_{\rm c}(T)}{I_{\rm c}(0)} = \frac{\Delta (T)}{\Delta (0)}\tanh{\left[\frac{\Delta (T)}{2k_{B}T}\right]},
\end{equation}
where $I_{\rm c}(0)$ and $\Delta (0)$ are the zero temperature critical current and superconducting gap respectively. 
  $\Delta(T)$ can be described by an interpolation approximation to the Bardeen-Cooper-Schrieffer (BCS) theory temperature-dependent superconducting energy gap, 
\begin{equation} \label{eq:BCS}
\Delta(T) = \Delta(0)\tanh{2.2\sqrt{(T_{c}-T)/T)}}.
\end{equation}
 The fit in Fig \ref{fig:Figure_3}b shows that the critical current of sample S1 is reasonably well described by AB theory, in agreement with previous reports on NbSe$_{2}$ Josephson junctions,\cite{Yabuki2016} despite the assumption of a symmetric single gap superconductor. This allows us to estimate $\Delta (0)$ = 0.90 mV and $I_{c}(0)$ = 1.06 mA.

 However, not all devices fabricated behave as underdamped junctions, and upon examination of a second device presented in Fig \ref{fig:Figure_3}c (S2, $\theta\sim$ 20$^{\circ}$), we observe $I$-$V$ characteristics which are non hysteretic, indicative of a junction which is in the strongly overdamped regime ($\beta_{c} \ll 1$).
 Here, non-linear features arise due to quasiparticle currents that reveal information about the superconducting density of states.
 We examine this in Fig \ref{fig:Figure_3}d by taking the numerical differential of the measured $I$-$V$ curves. 
 From this we observe an intricate non-BCS like gap structure, with peaks located at $\pm$1.79 mV at $T$ = 3.75 K.
 Furthermore, we observe a shoulder-like hump at lower energy ($\pm$1 mV), similar to that seen in tunnelling spectroscopy of NbSe$_{2}$ using van der Waals tunnel barriers.\cite{Khestanova2018,Dvir2018}
  Using the BCS approximation in Eqn \ref{eq:BCS}, we can extract the zero temperature superconducting gap $\Delta(0)$.
 Despite sub-gap features indicative of non BCS-like behaviour, the measured $\Delta(T)$ fits reasonably well with $\Delta(0)$ = 0.93 mV as shown in Fig \ref{fig:Figure_3}e.

\subsubsection{Angular dependence}

 \begin{figure}[t]
 \centering
  \includegraphics[trim={0cm 0cm 0cm 0cm}, width=1\linewidth,clip=true]{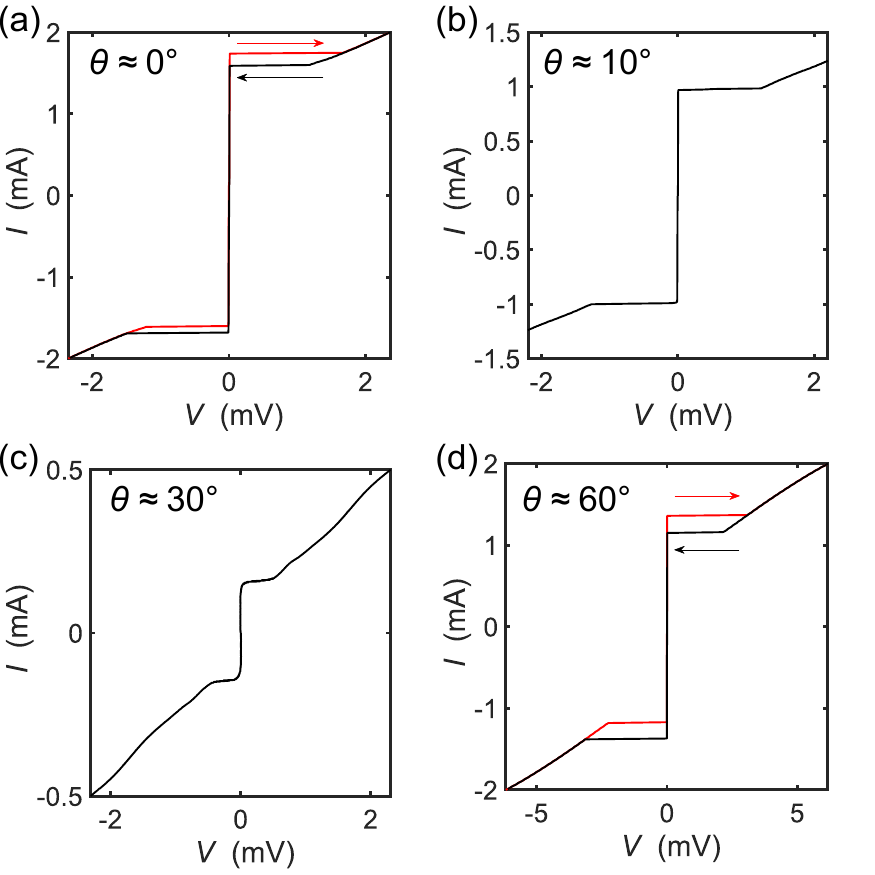}
  \caption{a) Current-voltage (\textit{I}-\textit{V}) characteristics for twisted NbSe$_{2}$-NbSe$_{2}$ junctions with twist angles $\theta$ $\approx$ 0$^{\circ}$, b) 10$^{\circ}$, c) 30$^{\circ}$, d) 60$^{\circ}$. The red and black arrows indicate the direction of the current sweep.}
  \label{fig:Figure_2}
\end{figure}

\begin{figure*}[t]
 \centering
  \includegraphics[trim={0cm 0cm 0cm 0cm}, width=0.75\linewidth,clip=true]{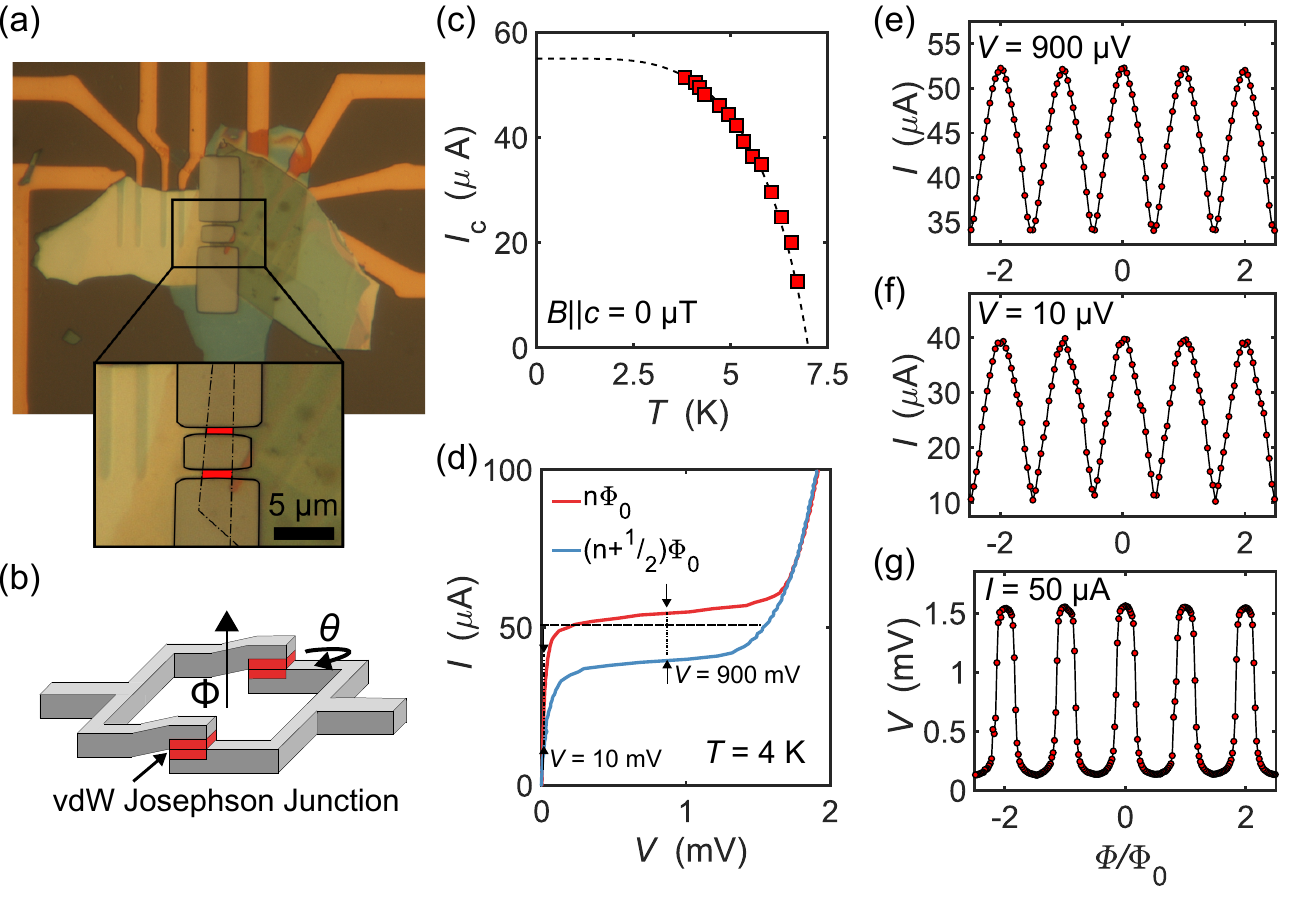}
   \caption{(a) Optical image of an etched SQUID structure. The insert shows a close up of the structure, with the overlap region between the two NbSe$_{2}$ flakes highlighted in red. (b) Schematic diagram of the SQUID device shown in (a). (c) SQUID critical current as a function of temperature in zero applied field. (d) Current-voltage ($I$-$V$) characteristics of a NbSe$_{2}$-NbSe$_{2}$ SQUID at 4 K. The red (blue) curve corresponds to the maximum (minimum) value of the positive critical current $I_{\rm c}$ within one period. (e) Modulation of $I_{c}$ as a function of the applied magnetic flux under a voltage bias of $V$ = 900 $\mu$V (f) and $V$ = 10 $\mu$V. (g) Voltage modulation as a function of the magnetic flux under a current bias of $I$ = 50 $\mu$A.
}
  \label{fig:Figure_4}
\end{figure*} 

Having observed strikingly different characteristics in NbSe$_{2}$-NbSe$_{2}$ junctions prepared using the same fabrication method, we turn our attention to the role of twist angle in these devices. 
 Using the method presented in Fig \ref{fig:Figure_1}a, we are able to fabricate devices with a determined misalignment angle respective to the crystallographic axes of the two flakes. 
The results are shown in Fig \ref{fig:Figure_2}a-d in which we present the current-voltage ($I$-$V$) characteristics of four NbSe$_{2}$-NbSe$_{2}$ junctions with twist angles in the range $\theta$ $\approx$ 0 - 60$^{\circ}$.
  First we examine the $\theta \approx$ 0$^{\circ}$ device, in which hysteresis is observed in the \textit{I}-\textit{V} characteristics, indicating the junction is an underdamped Josephson junction.
In samples with larger twist angles ($>$10$^{\circ}$, Fig \ref{fig:Figure_2}b), this hysteresis disappears, giving rise to reversible junctions which show features of quasiparticle gap structure at angles $> $20$^{\circ}$ (Fig \ref{fig:Figure_2}c).
 As the twist angle is increased beyond about 40$^{\circ}$, the gap features disappear and the hysteresis eventually returns near $\theta$ $\approx$ 60$^{\circ}$ (Fig \ref{fig:Figure_2}d).

Angle resolved photo-emission spectroscopy (ARPES) data\cite{Yokoya2001} and density functional theory calculations\cite{Johannes2006} show that there are five electronic bands crossing the Fermi energy in NbSe$_{2}$.
 Of these, one is a small Se-$4p$ $p_{z}$-derived "pancake"-shaped hole pocket, while the other four are Nb-$4d$-derived bands with roughly cylindrical Fermi surfaces centered at the $\Gamma$ and $K$ points in the Fermi surface.
 Based on ARPES measurements, the Se pancake has been shown to exhibit no superconducting gap, while the Nb-derived sheets display superconductivity which is strongly anisotropic in $k$.\cite{Borisenko2009,Rahn2012} 
 This anisotropy of the superconducting order parameter is characterised by maxima in the gap at 60$^{\circ}$ intervals around the Fermi sheets. 
 From this, we hypothesise that at twist angles close to 30$^{\circ}$, tunnelling is dominated by processes that couple regions of the Fermi surface with maximum gap in one layer with regions of minimum gap in the other layer.
This angular-dependent selectivity in the tunnelling process suppresses the critical current $I_{c}$, leading to a lower value of $\beta_{c}$, and non-hysteretic $I$-$V$ characteristics.

The observed twist dependence could also be related to the specific atomic arrangement at the interface which depends on the flake terminations as well as any effects arising from relative lateral displacements of the two flakes. However, the latter are expected to be negligible in our large overlap regions, which are many hundreds of unit cells wide.
We note that we have minimised any variation in the interface quality and homogeneity by performing all fabrication in the inert environment provided by a nitrogen glovebox.

\subsubsection{SQUID performance}

Having established the ability to fabricate high quality non-hysteretic Josephson junctions, we examine the ability to use them in more complex superconducting devices such as superconducting quantum interference devices (SQUIDs).
 Junction devices were patterned into a SQUID geometry using reactive ion etching as described in Methods.
 An image of a typical device is shown in Fig \ref{fig:Figure_4}a, and consists of a superconducting loop between two $\sim$600 nm wide Josephson junctions formed in the overlap region of the two NbSe$_{2}$ flakes.
 The SQUID loop is of width, $W$ $\approx$ 3.5 $\mu$m, and length, $L$ $\approx$ 7.0 $\mu$m, creating an internal hole of area $\approx$ 24.5 $\mu$m$^{2}$.
 When an external magnetic field $B\parallel_{c}$ is applied, phase shifts are induced between the two junctions leading to interference and a net critical current that oscillates with a period $\Phi_{0}/A$. In practice all real SQUIDs have a finite self-inductance that gives rise to screening currents that can often result in undesirably low values of current modulation $\Delta I_{c}$($B\parallel_{c}$).

 To investigate this in our NbSe$_{2}$-NbSe$_{2}$ SQUIDs, we used electrical transport measurements to characterise them at temperatures down to $T$ = 3.75 K.
 Figure \ref{fig:Figure_4}c shows the temperature dependence of the critical current. We observe no reduction of the critical temperature, indicating that etching caused no deterioration in the junction quality.
 Fig \ref{fig:Figure_4}d shows the current-voltage characteristics for a NbSe$_{2}$ SQUID, with the two curves corresponding to the maximum, $I_{c}^{max}$, and minimum, $I_{c}^{min}$, measured values of the critical current within a single oscillation period.
  The modulation of the critical current, $I_{c}$($\Phi$) for two different voltage set points, is shown in Fig \ref{fig:Figure_4}e,f where $\Phi$ is the applied magnetic flux.
  Under a bias voltage of $V$ = 900 $\mu$V the current modulation depth is 33$\%$, which increases to 75$\%$ for a bias of $V$ = 10 $\mu$V. The noise level of the signal increased slightly at $V$ = 10 $\mu$V due to small temperature oscillations arising from the refrigeration cycle of the closed-cycle cryostat.
The inductance parameter characterises the amplitude of the current modulation and is given by $\beta_{L} = 2\pi LI_{c}/\Phi_{0}$, where $\Phi_{0}$ is the superconducting flux quantum and $L$ is the inductance of the SQUID loop.
 Fitting our measured $\Delta I_{c}$ data to numerical models we estimate that $\beta_{L} \approx$ 2, and $L \approx$ 10 pH, which is consistent with other estimates of the self-inductance based on the SQUID geometry.
 The size of the SQUID loop $A$ = 24.5 $\mu$m$^{2}$ leads to a theoretical value of $\Phi$/$\Phi_{0}$ $\simeq$ 84 $\mu T$, which is close to the measured value of $\Phi/\Phi_{0}$ = 78 $\mu T$.
  Additionally, no decrease in the overall current modulation, $\Delta I_{c}$, was observed up to 20 mT, the maximum output of the electromagnetic coil.

An important property of modern SQUIDs is the realisation of non-hysteretic $I$-$V$ characteristics ($\beta_{c} \ll$ 1), allowing the use of flux-locked loop feedback schemes to dramatically increase the flux resolution.\cite{Kleiner2004, Martinez-Perez2017}
 We are able to tune this desirable property in our NbSe$_{2}$ SQUIDS by deterministically controlling the twist angle during device fabrication, with a wide range of angles over which $\beta_{c}$ is found to be sufficiently small.
 Figure \ref{fig:Figure_4}g shows the voltage under a bias current of 50 $\mu$A, as a function of magnetic field applied to the SQUID, revealing large voltage modulations with a depth of $\approx$ 1.4 mV.
 The voltage modulation depth is characterised by $\Delta V \simeq \delta R \Delta I_{c}$, where $\Delta I_{c}$ is the critical current modulation depth, and $\delta$R is the differential resistance of the SQUID calculated here to be $\delta$R $\simeq$ 28 $\Omega$.
 A large value of $\Delta V$ is required to achieve low flux noise SQUID devices as it minimises the noise contribution from the input amplifier $S^{1/2}_{V,a}$ to the white flux noise, given by $S^{1/2}_{\Phi,a} \simeq \Phi_{0} S^{1/2}_{V,a} / \pi \Delta V$, in flux-locked operation.\cite{Trabaldo2019}
No degradation in $T_{c}$ or $I_{c}$ of these SQUIDs was observed after multiple cooling cycles or when remeasured after storage in a nitrogen glovebox for 4 weeks, indicating that they are both stable and suitable for repeated long term use.

\subsection{Discussion}

In conclusion, we have developed a method to fabricate Josephson junctions and SQUIDs based on a twisted van der Waal heterostructure architecture.
 The formation of Josephson junctions is achieved using a dry transfer method and requires no 'wet' lithographic steps. 
The result is a junction device whose $I$-$V$ characteristics can be deterministically tuned via the twist angle.
 A single lithographic process is then implemented to shape the Josephson junction into a SQUID geometry with typical loop areas of $\simeq$ 25 $\mu m^{2}$ and weak links $\simeq$ 600 nm wide.
 We obtain voltage modulation depths of $\Delta$V $\simeq$ 1.4 mV and current modulation depths of $\Delta$I $\simeq$ 75$\%$ from devices which display long term stability.

 Our method demonstrates the ability to fully integrate 2D materials into the design of a high-performance SQUID, paving the way for designer superconducting circuits through incorporation of other 2D materials, each with distinct electronic properties.
The utilisation of van der Waals bonded circuits may also have other advantages: here the devices are only anchored to the substrate by the weak vdW force, allowing the possible pick-up and transfer of characterised SQUIDs onto other substrates or materials, while the intrinsically thin nature of 2D materials may allow their incorporation into flexible electronic circuits.\cite{Kim2015} 
 Other applications for the technology is in the design of superconducting qubits such as those based on flux qubit architectures.\cite{Chiorescu2004,Yan2016} It is well known that the performance of these is often limited by dissipation, decoherence and noise from two-level defect systems. The single crystalline structure of 2D flakes, along with low defect densities, may provide circuit components with superior performance.

\subsection{Method.}
  Thin flakes of NbSe$_{2}$ were mechanically exfoliated from high quality single crystals onto silicone elastomer polydimethylsiloxane (PDMS) stamps.
 Flakes of suitable geometry and thickness were then sequentially transferred onto Si/SiO2 (300 nm oxide) substrates with pre-patterned Au contacts.
 To ensure crystallographic alignment of the twisted vdW junctions, one single exfoliation of the bulk single crystal is performed which results in multiple thin flakes with aligned crystallographic axes.
  From this, two neighbouring flakes were chosen with a rotation of the substrate performed after stamping the first flake.
  The resulting device is thus a twisted vdW junction with the two flakes misaligned by a chosen angle with $\pm$1$^{\circ}$ accuracy.
 The entire dry transfer set-up is housed in a nitrogen glovebox with an oxygen and moisture content <1 ppm, ensuring an oxide free interface.
  Devices selected for fabrication into SQUIDs were subsequently capped with a thin layer ($\sim$20 nm) of hexagonal boron nitride (h-BN) using the same dry transfer technique.
 Following this, a polymethyl methacrylate e-beam resist was spin coated onto the substrate, which was then kept under vacuum (10$^{-6}$ mbar) overnight to remove any solvent.
 Next, standard e-beam lithography (EBL) was used to define the area of the junction to be etched to form the SQUID loop, before transfer into an inductively coupled plasma etcher.
 The sample was then reactively ion etched using an O$_{2}$ + SF$_{6}$ mixture, after which the sample was immersed in acetone to remove leftover resist before storage in a nitrogen glovebox.

\subsection{Associated Content}
\subsubsection{Supporting Information}
The Supporting Information is available at xxx.

\subsection{Author Information}

\subsubsection{Corresponding Author}
*E-mail: L.S.Farrar@bath.ac.uk

\subsubsection{Present Address}
$^{\dagger}$Department of Physics, University of Bath, Bath BA2 7AY, United Kingdom

\subsubsection{Acknowledgements}
The authors thank Peter Heard at the IAC, University of Bristol, for their useful discussions and supporting work at the early phases of this study. The research was funded by the Bath/Bristol Centre for Doctoral Training in Condensed Matter Physics, under the EPSRC (UK) Grant No. EP/L015544. The work at the University of Warwick is supported through EPSRC Grant EP/T005963/1. A.N, Z.J.L, and S.D acknowledge financial support from the Royal Society.

\subsubsection{Author contributions}
L.F fabricated the devices under the supervision of S.B, with support from A.N. and Z.J.L under the supervision of S.D. G.B grew the single crystals. L.F performed the measurements. L.F and S.B wrote the paper with comments and discussion from all authors.

\end{document}


\oldmaketitle

\setcounter{equation}{0}
\setcounter{figure}{0}
\setcounter{table}{0}
\setcounter{page}{1}
\makeatletter
\renewcommand{\theequation}{S\arabic{equation}}
\renewcommand{\thefigure}{S\arabic{figure}}
\renewcommand{\bibnumfmt}[1]{[S#1]}
\renewcommand{\citenumfont}[1]{S#1}

\begin{figure*}[t]
 \centering
  \includegraphics[trim={0cm 0cm 0cm 0cm}, width=1\linewidth,clip=true]{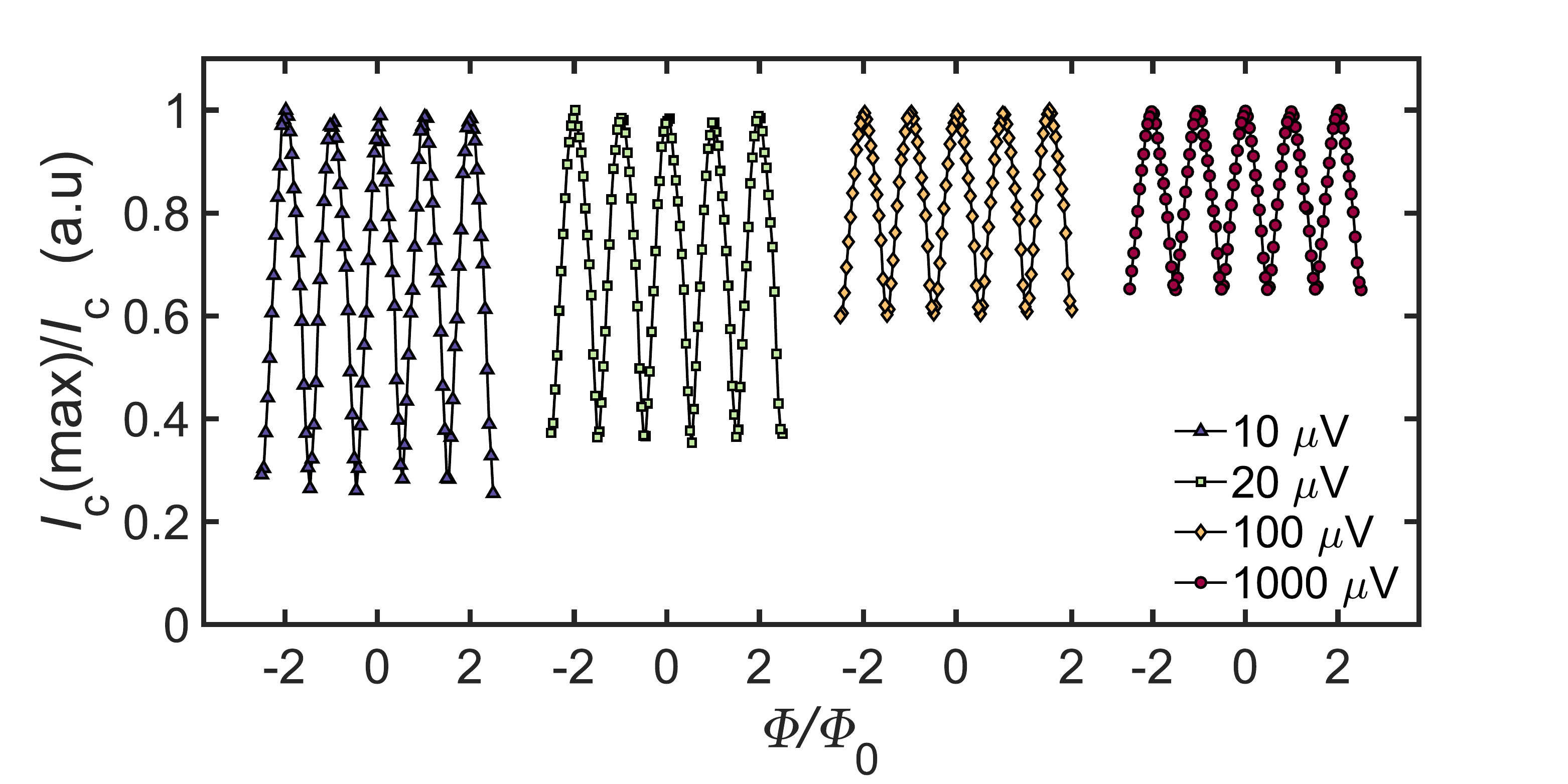}
  \caption{
Modulation of $I_{\mathrm{c}}$ of an NbSe$_{2}$-NbSe$_{2}$ SQUID at $T$ = 4 K as a function of the applied magnetic flux under various bias voltages. 
   }

  \label{fig:SI_Figure_1}
\end{figure*}

\begin{figure*}[t]
 \centering
  \includegraphics[trim={0cm 0cm 0cm 0cm}, width=0.65\linewidth,clip=true]{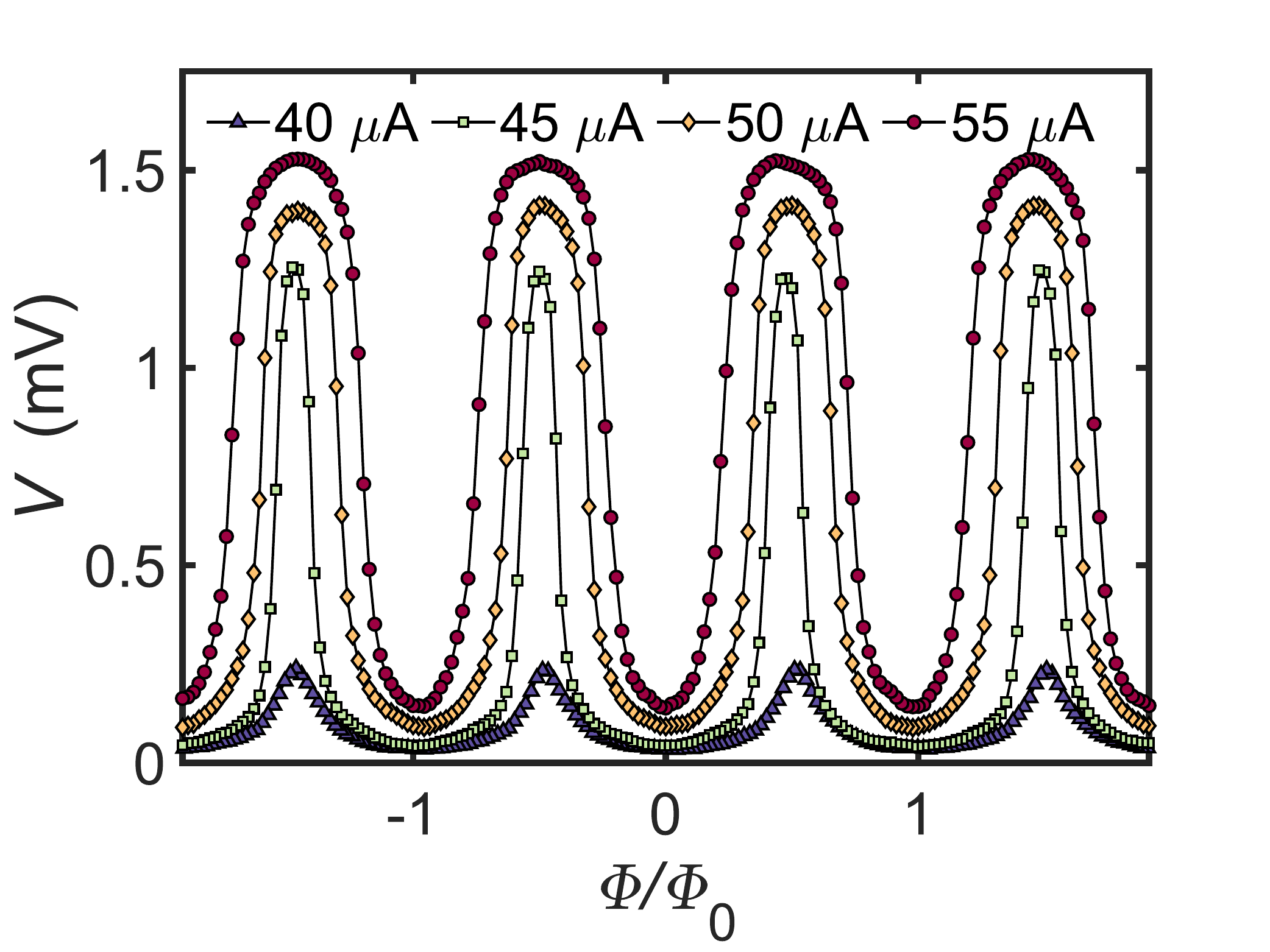}
  \caption{
Voltage modulation of an NbSe$_{2}$-NbSe$_{2}$ SQUID at $T$ = 4 K as a function of the magnetic flux under various bias currents.
   }

  \label{fig:SI_Figure_2}
\end{figure*}